# Enhanced Algorithm for Link to System level Interface Mapping

Shahid Mumtaz
Institute of Telecommunication
Aveiro,Portugal
smumtaz@av.it.pt

Alitio Gamerio &Raool Sadeghi
Institute of Telecommunication
Aveiro,Portugal
agm@av.it.pt

*Abstract*—**The current SINR mechanism does not provide the base station (BS) with any knowledge on the frequency selectivity of channel from mobile service station(MSS). This knowledge is important since, contrary to the AWGN channel, in a frequency selective channel there is no longer a 1 to 1 relation between amount of increase in power and amount of improvement in "effective SINR" 1. Furthermore, the relation is dependent on MCS level. This lack of knowledge in the BS side results in larger fade margins, which translates directly to reduction in capacity.**
**In this paper we propose a enhanced algorithm on the EESM model with the weighted beta ($\beta$) that provides the BS with sufficient knowledge on the channel-dependent relationship between power increase, MCS change and improvement in effective SINR**.

### INTRODUCTION

A great deal can be learned about an air interface technology by analyzing its performance in a link level setting consisting of one base station and one mobile station. This link level analysis is of fundamental importance for the evaluation of the technologies associated to the given air interface, namely for the study of the variation of the Bit Error Rate (BER) with the Signal to Noise Ratio (SNR) per bit sent along the transmission chain, under the influence of such an aggressive medium for signal transmission as the wireless mobile channel. In real-world, where multiple base stations are deployed in a service area and operating in the presence of a large number of active mobile users, the system performance can only be evaluated through a system-level analysis, where the point-to-point radio link communication scenario is replaced by one in which all radio links among the mobile and base stations must be considered. Typically, network simulations are divided into two parts: link and system level simulations. Although a single simulator approach would be preferred, the complexity of such simulator (including everything from transmitted waveforms to multi-cell network) is far too high with the required simulation resolutions and simulation times.
Therefore, separate link and system level simulations are needed. Typically, the link level simulator is able to predict the receiver Frame Erasure Rate/Bit Error Rate (FER/BER) performance, taking into account channel estimation, interleaving and decoding and is needed to build a model for the system level simulator, which is needed to model a system with a large number of mobile and base stations and the algorithms operating in such a system.
In system level simulations, we will focus on making transmission adaptations to optimize system performance and get better understanding of the user performance in various deployment scenarios. For complexity reasons system level evaluations have to rely on simplified Physical (PHY) -layer models that still must be accurate enough to capture the essential behavior. So, the modeling method of link layer is very essential and important.
The Block error rate (*BLER*) performance versus signal to inference and noise ratio (*SINR*) averaged over all channel realizations of one specific channel model has been widely used as the interface between the PHY- and system-level simulators. But in many cases, the specific channel realization encountered may perform significantly different from the average performance. Consequently, many novel modeling approaches accounting for the instantaneous channel and interference conditions are introduced such as CESM (Capacity based effective SINR Mapping), EESM (Exponential effective SNIR Mapping ) and MIESM (Mutual Information based effective SINR Mapping) have been brought forward.
The abstraction method adopted in this paper is based on the EESM algorithm over link layer of 802.16e system and some modifications have been made to improve the performance prediction accuracy. This paper is organized as follow. Section II explain the simulation scenario using in this paper, Section III explains the Link layer Abstraction based on EESM, Section IV presented the Enhanced Algorithms based on EESM and conclusion is presented in section V.

### II SIMULATION SCENARIO

Link level simulation chain is shown in Figure 1 which is written in System C and run under Linux platform with gcc complier 1.95 or more.
Randomization is a process to systematically or randomly reorder or randomize the transmitted data. It is employed to minimize the possibility of transmission of an un-modulated carrier and to ensure adequate numbers of bit transitions to support recovery. Randomization is achieved by XORing the data blocks with a pseudo-random binary sequence (PRBS) generated using a certain polynomial [1]. Another purpose of randomization is to encrypt the transmitted data blocks to prevent any unintended receiver from decoding the data.

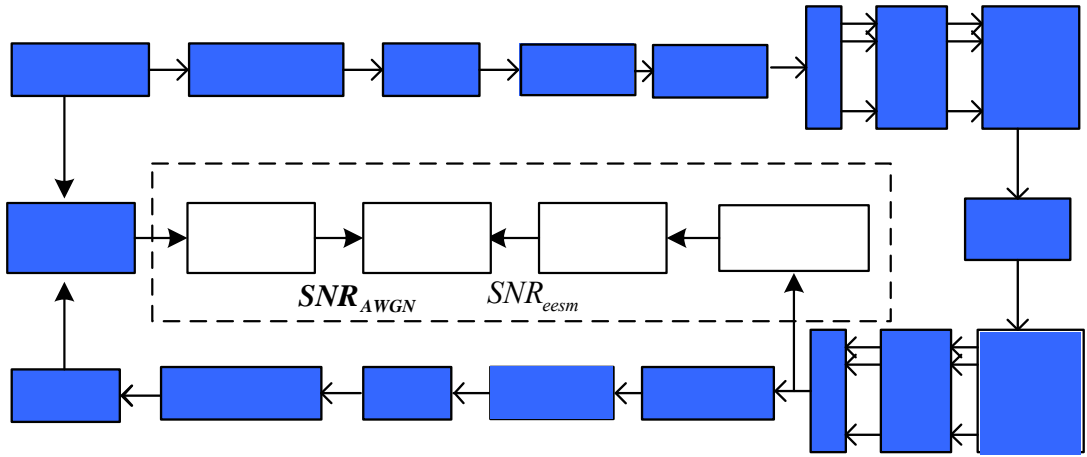

Figure 1:Link Level Simulation Chain

In encoding FEC (Forward Error Correction) process is used to maximize the possibility of detecting and possibly recovering the corrupted received data by adding redundancy to the transmitted data. WiMAX-OFDM standard specifies three methods of FEC: Reed-Solomon concatenated with convolution coding (RS-CC), block turbo coding (BTC), and convolution turbo coding (CTC). WiMAX-OFDMA specifies five methods of channel coding: convolution coding (CC) with tail biting, block turbo coding (BTC), convolution turbo coding (CTC), low density parity check coding (LDPCC), and CC with zero tailing. The most common channel coding method is CTC. The encoded data from the previous step go through a two-step process. The first step ensures that adjacent encoded bits are mapped into non adjacent subcarriers to provide frequency diversity and to improve the performance of the decoder. The second step maps the adjacent bits to the less and more significant bits of the constellation. The modulation of data bits depends on the modulation scheme used. WiMAX takes into consideration channel quality to choose the correct modulation scheme. The modulation scheme is selected per subscriber to achieve the best performance possible. The number of bits per symbol (time) depends on the modulation scheme used, for QPSK (Quadrature Phase Shifting Keying) it is 2, for 16- QAM (Quadrature Amplitude Modulation) it is 4, and for 64-QAM it is 6.Once the signal has been coded, it enters the modulation block. All wireless communication systems use a modulation scheme to map coded bits to a form that can be effectively transmitted over the communication channel.Thus, the bits are mapped to a subcarrier amplitude and phase, which is represented by a complex in-phase and quadrature-phase (IQ) vector. WiMAX specifications for FFT OFDM PHY layer define three types of subcarriers; data, pilot and null, as shown in Figure 2.
Each OFDM symbol is composed of data subcarriers, zero DC subcarrier, pilot subcarriers, and guard carriers according to permutation schemes used. For example table 1 shows the downlink PUSC configuration. Furthermore, preambles consisting of training sequences area appended at the beginning of each burst. These training sequences are used for performing an estimation of the channel coefficients at the receiver. The signal is converted to the time domain by means of the inverse fast Fourier transform (IFFT) algorithm, and finally, a cyclic prefix (CP) with the aim of preventing inter-symbol interference is added.

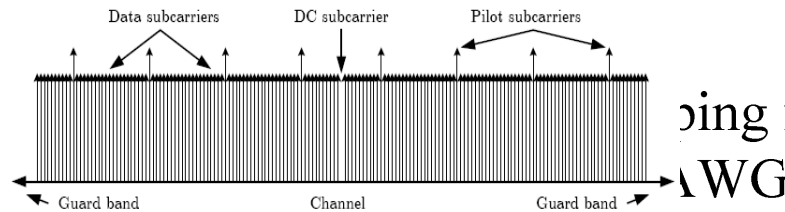

Figure 2: OFDM Symbol Structure in Time Domain

### III   LINK LAYER ABSTRACTION

In order to simulate WiMAX, we need to simulate 1024 (or more) subcarriers, the effect on noise on each of these subcarriers and their effect on the received FEC blocks. Such a simulation can be very complex and time consuming. This complexity can be avoided by modeling the channel as an additive white Gaussian noise (AWGN) channel with a single effective SINR. Wireless scientists have developed several ways to combine the SINRs of multiple subcarriers into a single effective SINR. One of the commonly used methods is the so called "Exponential Effective SINR Mapping" or EESM.EESM is used to map the instantaneous values of SINRs to the corresponding BLER (Block Error Rate) value.

Although EESM was introduced to work with SIR (Signal to Interference Ratio), it works with SNR as well.
EESM is a simple mapping method used when all the subcarriers of a specific subscriber are modulated using the same Modulation and Coding Scheme (MCS) level. The basic idea of EESM is to find a compression function that maps the set of SINRs. to a single value that is a good predictor of the actual BLER [2]. Figure 3 shows the main purpose behind

| Parameter Name | Value |
|---|---|
| System Channel Bandwidth (MHz) | 10 |
| Sampling Frequency ($F_P$ in MHz) | 11.2 |
| Subcarrier Frequency Spacing ($f$ kHz) | 10.94 |
| FFT Size ($N_{FFT}$) | 1024 |
| UP/DL | DL |
| Null Subcarriers | 184 |
| Pilot Subcarriers | 120 |
| Data Subcarriers | 720 |
| Data Subcarriers per Subchannel | 24 |
| Number of Subchannels ($N_s$) | 30 |
| | 91.4 |
| Useful Symbol Time ($T_b=1/f$) in µs | |
| | 11.4 |
| Guard Time ($T_g=T_b/8$) in µs | |

using EESM function. Here, BLER refers to block error rate and PER refers to packet error rate. Note that average SINR is not a good predictor of actual BLER or PER (Packet Error Rate).

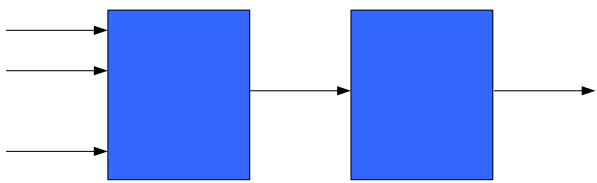

Figure 3: SINR Compression

EESM is a channel-dependent formula that maps power level as well as MCS level to SINR values in the AWGN (Additive White Gaussian Noise) channel domain. Such function allows its mapping along with AWGN assumptions (such as effect of increase in power, CINR/MCS threshold tables) to predict the effect of MCS and boosting modification. The method has been shown to yield an accurate estimation of the AWGN-equivalent SINR (henceforth referred to as .effective SINR.) for frequency selective channels [3].
In case of multi-carrier transmission as in WiMAX, the set of subcarrier SINRs are mapped with the help of EESM formula into a scalar instantaneous effective SINR value. An estimate of the BLER value is then obtained, using the effective SINR value, from basic AWGN link-level performance. The mapping of the effective SINR value to the corresponding BLER value will use either a look-up table for the mapping function or use an approximate analytical expression if available. The EESM method estimates the effective SINR using the following formula

$$\gamma_{eff} = EESM\ (\gamma, \beta) = -\beta \ln \frac{1}{N} \sum_{i=1}^{N} e^{\frac{-\gamma_i}{\beta}} \qquad (1)$$

Where, $\gamma$ is a vector [$\gamma_I$, $\gamma_2$., $\gamma_N$ ] of the per-subcarrier SINR values, which are typically different in a frequency selective channel. $\beta$ is the parameter to be determined for each Modulation Coding Scheme (MCS) level and $N$ is number of data subcarrier(720 in case of downlink PUSC) and this value is used to adjust EESM function to compensate the difference between the actual BLER and the predicted BLER.
To obtain $\beta$ value, several realizations of the channel have to be conducted using a given channel model (e.g., Pedestrian (Ped B) and Vehicular (Veh A)). Then BLER for each channel realization is determined using the simulation. Using the AWGN reference curves generated for each MCS level, BLER values of each MCS is mapped to an AWGN equivalent SINR. These AWGN SINRs for n realizations can be represented by an n-element vector SINR$_{AWGN}$. Using a particular $\beta$ value and the vector $\gamma$ of subcarrier SINRs, an effective SINR is computed for each realization. For n realizations, we get a vector of computed effective SINRs denoted by SINR$_{EESM}$. The goal is to find the best possible $\beta$ value that minimizes the difference between computed and actual effective SINRs:

$$\beta = \arg \min_{\beta} \left\| SNR_{AWGN} - SINR_{eesm} (\beta) \right\| \qquad (2)$$

The four steps to obtain beta value are as follows. First, generate an AWGN curve for a specific MCS level. Second, measure the SINR per tone (subcarrier) values for the same MCS level using the desired channel model (for instance Ped A or Ped B). Many channel realizations are required and SINR per tone values should be converted to one scalar value to represent the channel SINR using EESM formula. The third step is to compare the two values gained from the previous steps (SINR$_{EESM}$ and SINR$_{AWGN}$). The comparisons for many SNR-pair values will yield a mean squared difference for a given beta value The beta value that gives the minimum difference is selected as the optimal value. In the first step the AWGN channel model is used to generate the reference curve. The BLER values that are of interest are those that result in a satisfactory operation. This range includes small BLER values close to zero. Figure 4 shows an example of AWGN reference curve generated by the simulation process for QPSK with coding rate 1/2,2/3,3/4.
The second step is to get the SINR per tone values. First, all data subcarriers SNR values are stored for a single realization. Then a set of Gaussian random numbers with length equals to the number of the number of data subcarriers is generated. The sum of the both sets represents the SINR per tone values. To get one scalar value that represents the channel SINR (or effective SINR), EESM formula is used as shown in equation 1. Since EESM SINR value depends in the chosen beta value. The calibration process using equation 2 formula is used to minimize the difference between the expected and the simulated value of SNR.

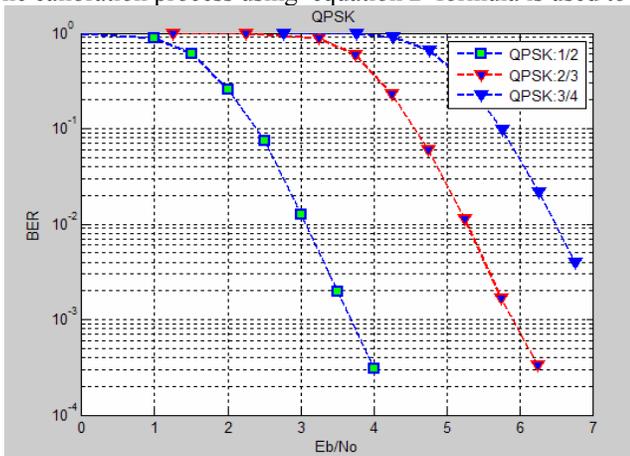

Figure 4: AWGN reference cruve for QSPK

Beta values of different format are trained on PB and VA channel respectively through adequate link layer simulation of 802.16e system. The obtained beta values for look up are shown in the following Table 2 and Table 3.

Simulations are done using SISO channel, PUSC mode, SCM channel model with the velocity of 3Km/h & 60Km/h,100 independent channel realizations with CTC and Ideal channel estimated is assumed. Beta values trained for PB and VA channel are quite similar in most cases, coinciding with the theory that the beta training should be independent of channel realizations. There are some differences when the higher order modulation is adopted, therefore, two beta tables are presented for different models in order to guarantee higher reliability of abstraction especially for higher order modulation. Figure 5 and Figure 6 shows the beta training for VA and PB channel

## IV    Enhanced Algorithms

The current 802.16e SINR reporting mechanism requires the MSS(Mobile station) to report a straightforward CINR(Carrier to Interference plus Noise Ratio) measurement. This mechanism does not provide the BS with any knowledge on the frequency selectivity of the channel and noise (especially prominent with partially loaded cells and with multipath). This knowledge is important since:

• Two channel realizations with the same average CINR may cause substantially different frame error rate (FER) depending on the instantaneous channel variation. Without a proper metric to reflect the channel realization, the base station is unable to provide accurate link adaptation.

• Contrary to the AWGN channel, in a frequency selective channel there is no longer a 1 to 1 relation between amount of increase in power and amount of improvement in "effective SINR". Furthermore, the relation is dependent on the modulation and coding scheme (MCS) level. This lack of knowledge in the BS side results in larger fade margins. Thus the current channel quality report scheme would lead to reduction in system capacity.

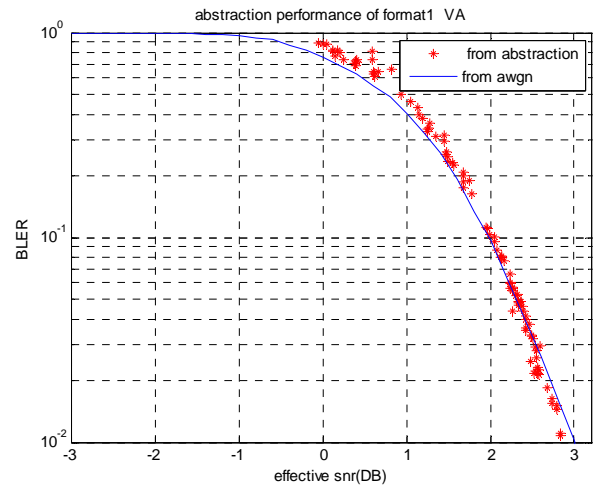

Figure 5: Predicted BLER VS. Simulated BLER

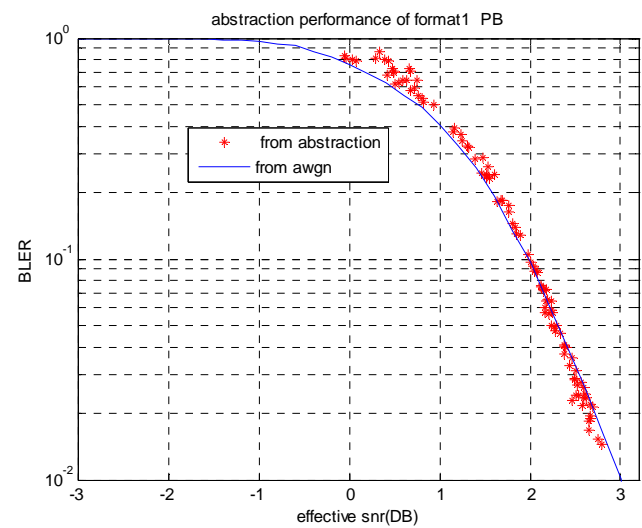

Figure 6: Predicted BLER VS. Simulated BLER

In general, we would like the MSS to report the effective SINR to the BS, and have the BS decide what modulation and coding to use and with what power boosting. This is complicated by the fact that the relationship between increase in power and increase in effective SINR is both channel-dependent and MCS-dependent. In context of EESM, this implies that for each MCS a different $\beta$ should be utilized, and for each such $\beta$, different boosting should be considered.

It is well known that the influences of *SNR* distance on the *PER* performance varies a lot in different parts of the performance curve. And when *SNR* value gets higher from a low start point, *PER* performance is more sensitive to the *SNR* difference.

The square difference between $SNR_{eesm}$ and the effective $SNR$ could be weighted by the relative difference between the current effective $SNR$ value and the $SNR$ value when the BLER begins to drop on the AWGN performance curve, which highlights the influences of the $SNR$ differences in the high $SNR$ region on the $PER$ performance. The cost function needed to be minimized is expressed in equation 3 , where $\mathbf{SNR_{eff}}$ and $\mathbf{SNR_{eesm}}$ are vectors with the size of simulated channel realization number, and $\mathbf{W}$ is the weight vector with the same size and is expressed in equation . $SNR_{start}$ is the SNR point where the BLER begins to drop from 1 ( BLER=0.99 is assumed) on the AWGN performance curve. The minimization algorithm is implemented based on golden section search and parabolic interpolation.

$$F(\beta) = \left\| SNR_{eff} - SNR_{eesm}(\beta)\Delta W \right\|^2 \qquad (3)$$

$$W = (\frac{SNR_{eff} - SNR_{start}}{SNR_{start}})^2 \qquad (4)$$

As a result, the BS is required to know the dependence of effective SINR on weighted β and power increase; thus computation of equivalent SNR can no longer remain solely in the MSS's territory.

The increase of $\gamma_{eff}$ due to boosting in weighted β dependent, where $\mathbf{B_\varphi}$ denotes the weighted boost ratio.

$$EESM(\gamma.B_\varphi, \beta\Delta W) = -\Delta\beta.\ln(\frac{1}{N}\sum_{i=1}^{N} e^{\frac{\gamma_i.B_\varphi}{\beta\Delta W}}) \qquad (5)$$

$$\neq EESM(\gamma.\beta\Delta W).B_\varphi$$

This implies that EESM is a two-dimensional mapping of weight boost level and an MCS dependent quantity (weighted β ) to effective SINR. However, we can simplify by observing that which shows that given an SINR-per-tone vector it is sufficient for the BS to know the MSS-specific curve relating EESM to weighted β . Both boosting and rate adaptation can be done based on the same curve, thus reducing the mapping problem to one dimension.

We plot EESM as function of weighted β , for different cases. The first graph plots EESM for 4 different γ vectors, drawn from 24 independent Rayleigh distributions.

Both EESM and weighted β are plotted in dB. It can be seen that the graphs can be approximated locally as linear (in dB=>dB), and have overall a linear shape with saturation at weighted β >15dB. Saturation occurs for practically unachievable weighted β values. This linear shape may be used for compressing the curve for transmission to the BS.

For the purpose of fast MCS adaptation or Hybrid ARQ, the MSS needs to provide instantaneous effective SINR and BS may decide MCS and boosting, according to MSS instantaneous effective SINR. However the number of relevant rates is limited and their weighted β values are close. Furthermore, the boosting range is limited, so we are typically interested in a narrow region of the weighted β axis. Thus a local linear approximation suffices, and the graph may be compressed effectively. This implies one straightforward

$$EESM(B_\varphi, \gamma, \beta\Delta W) = -\Delta\beta.\ln(\frac{1}{N}\sum_{i=1}^{N} e^{\frac{\gamma_i.B_\varphi}{\beta\Delta W}})$$

$$= B_\varphi.(\frac{-\beta\Delta W}{B_\varphi}).\ln(\frac{1}{N}\sum_{i=1}^{N} e^{\frac{\gamma_i}{\beta\Delta W / B_\varphi}}) \qquad (6)$$

$$= B_\varphi.EESM(\gamma, \beta\Delta W / B_\varphi)$$

solution – the MSS can initially (e.g. on handover to a new cell) send a table of EESM SINR thresholds and β values for each MCS, and then at a higher speed transmit a local linear approximation for the EESM(β ) curve.

The accuracy of the EESM modeling technique as a predictor for the AWGN equivalent SINR was analyzed extensively for OFDM in [4][5][6]. In addition, we performed a short examination in order to validate the accuracy of EESM for 802.16. The following methodology was used. First, optimal β values were estimated for each MCS level. Then, the accuracy of EESM was evaluated:

The following figures 7,8,9 and 10 show, for each MCS (QPSK. 16-QAM), the distribution of the EESM fit error (on the left) and the mean SINR vs. EESM prediction error (on the right) for the channel realizations .

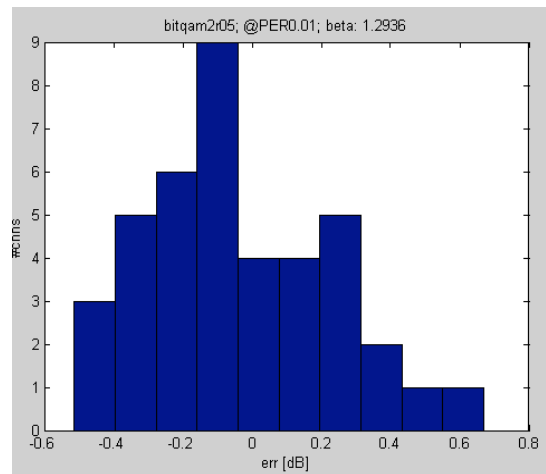

Figure 7: QPSK EESM fit error

The proposed mechanism is as follows:
a. MSS computes SINR-per-tone vectors for the purpose of EESM.

b.MSS computes the curve parameters of EESM(β) in the weighted β

 The range of interest depends on current MCS level, for example, an MSS that operates in the QPSK area should compute the local slope for the QPSK range of weighted βs rather than the local slope for the QAM-64 range of β s.

c. MSS sends the curve parameters to the BS, and updates the BS whenever these parameter change (due to change in channel conditions) – slow update.

MSS uses β values from a table of β per MCS (provided by the BS) to compute CINR measurement based on the EESM formula. These measurements are averaged.

d. The MSS compensates for implementation losses so that the transmitted CINR values are aligned with normalized threshold levels supplied by the BS.

e. A CINR report consists of a single CINR value. The MSS sends the CINR measurement that corresponds to one of the βs; this weighted β is selected using a rule, which ensures that the BS knows its value.

The BS now has all needed information (EESM CINR value, β for which it was computed, local-linear approximation of EESM (β)) in order to predict the effect of boosting and change of MCS level with the MSS's current channel conditions.

## V. CONCLUSIONS

 In system level simulations, we will focus on making transmission adaptations to optimize system performance and get better understanding of the user performance in various deployment scenarios. For complexity reasons system level evaluations have to rely on simplified Physical (PHY)-layer models that still must be accurate enough to capture the essential behavior. So, the modeling method of link layer is very essential and important. In this paper we presented the enhanced EESM modeling method that can be used for accurate link adaptation and accurate power boosting. The method provides the BS with sufficient knowledge on the channel-dependent relationship between MCS, power increase, effective SINR and describe in detail how to calculate beta.

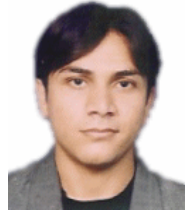

Shahid Mumtaz received his Masters degree in Electrical engineering from the Blekinge Institute of Technology from Sweden, Karlskrona , in 2005. He is working as Research Engineer at the Instituto de Telecomunicações, Pólo de Aveiro Portugal. His research interests include QoS in 3G/4G Networks, Radio Resource Management for wireless systems. His current reserach activities involve Cross-Layer Based Dynamic Radio Resource Allocation for WANs.

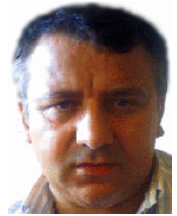

Atilio Gameiro received his Licenciatura (five years course) and his PhD from the University of Aveiro in 1985 and 1993 respectively. He is currently a Professor in the Department of Electronics and Telecommunications of the University of Aveiro, and a researcher at the Instituto de Telecomunicações - Pólo de Aveiro, where he is head of group. His main interests lie in signal processing techniques for digital communications and communication protocols.

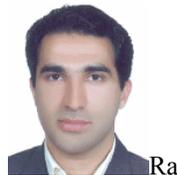

Rasool Sadeghi received his MSc. in 2004 in Telecommunication Engineering from Shiraz University, Iran. Then he joined to ITMC (Siemens partner in Iran) and worked on the project of TMN (Telecommunication Management Networks).In 2006-2007, he worked for TCE as a Network and Switching Engineer of GSM.. Since December 2007, he is a PhD student in Institute of Telecommunications at Aveiro university, Portugal. His research interests are network and radio resource management algorithms for wireless systems and cooperative diversity.

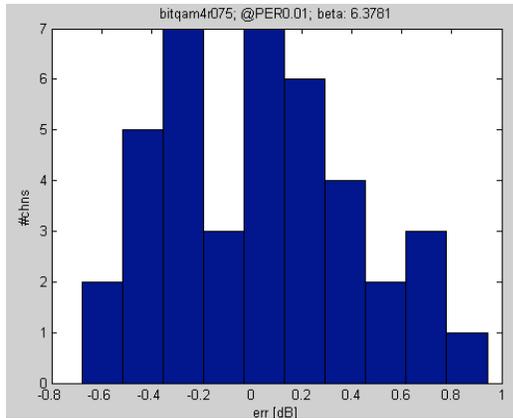

Figure 9: QAM16 EESM fit error

Figure 8: QPSK Mean SINR and prediction error per channel realization

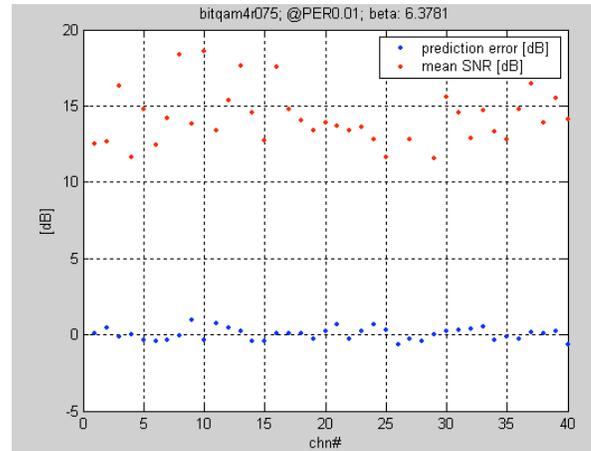

Figure 10 : QAM16 Mean SINR and prediction error per channel realization

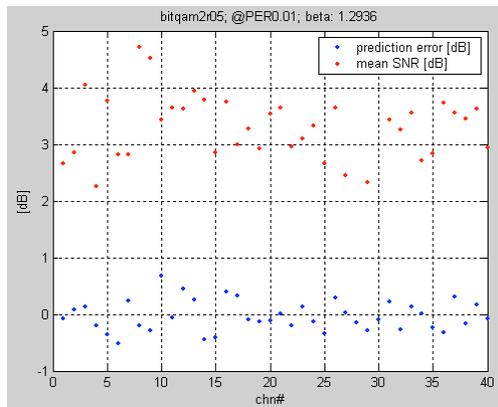

Table II  Beta values for PB channel (3Km/h)

| Format | 1 | 2 | 3 | 4 | 5 | 6 | 7 | 8 | 9 | 10 | 11 | 12 | 13 | 14 | 15 | 16 |
|--------|------|------|------|------|------|------|------|-------|-------|-------|-------|-------|-------|-------|-------|-------|
| Beta(dB) | 2.46 | 2.28 | 2.27 | 2.18 | 2.05 | 2.00 | 2.03 | 2.04 | 1.98 | 2.56 | 2.43 | 2.46 | 2.41 | 2.41 | 2.38 | 7.45 |
| Format | 17 | 18 | 19 | 20 | 21 | 22 | 23 | 24 | 25 | 26 | 27 | 28 | 29 | 30 | 31 | 32 |
| Beta(dB) | 7.14 | 7.00 | 7.34 | 6.89 | 8.93 | 8.87 | 8.85 | 11.31 | 11.11 | 11.09 | 13.80 | 13.69 | 14.71 | 14.59 | 15.32 | 15.29 |

Table III Beta values for VA channel (60Km/h)

| Format | 1 | 2 | 3 | 4 | 5 | 6 | 7 | 8 | 9 | 10 | 11 | 12 | 13 | 14 | 15 | 16 |
|--------|------|------|------|------|------|------|------|-------|-------|-------|-------|-------|-------|-------|-------|-------|
| Beta(dB) | 2.54 | 2.26 | 2.26 | 2.12 | 2.07 | 2.06 | 2.02 | 2.01 | 2.01 | 2.50 | 2.43 | 2.44 | 2.39 | 2.41 | 2.37 | 7.48 |
| Format | 17 | 18 | 19 | 20 | 21 | 22 | 23 | 24 | 25 | 26 | 27 | 28 | 29 | 30 | 31 | 32 |
| Beta(dB) | 7.14 | 6.92 | 7.53 | 6.82 | 8.93 | 8.87 | 8.90 | 11.43 | 11.16 | 11.01 | 13.74 | 13.70 | 14.68 | 14.55 | 15.17 | 15.27 |